\documentclass[twocolumn]{aa}
\usepackage{graphicx}
\usepackage{amsmath}
\usepackage{txfonts}

\newcommand{\multi}[3]{\ensuremath{\{#2\}\overset{#1}{=}\{#3\}}}
\newcommand{\sinc}{\ensuremath{\operatorname{sinc}}}
\newcommand{\Sum}{\ensuremath{\operatorname{Sum}}}
\newcommand{\tsum}{\ensuremath{\textstyle\sum}}
\newcommand{\set}[1]{\ensuremath{\{#1\}}}

\usepackage{natbib}
\begin{document}

\title{The achromatic chessboard,  a new concept of phase shifter for Nulling Interferometry}
\subtitle{I.  theory}
\author{Daniel Rouan \inst{1},  Didier Pelat \inst{2}}

\offprints{D. Rouan \email{daniel.rouan@obspm.fr}}

   \institute{ LESIA, Observatoire de Paris, CNRS, UPMC, Universit\'e Paris Diderot; 5 place Jules Janssen, F-92190 Meudon\\
              \email{daniel.rouan@obspm.fr}\\
 LUTH, Observatoire de Paris, CNRS, Universit\'e Paris Diderot; F- 92190 Meudon, \\
               \email{didier.pelat@obspm.fr}}
  \date{Received; accepted : 21 Feb 2008}

\abstract
% Context
{
Direct detection of a planet  around a star and its characterization for identification of bio-tracers in the mid-IR, requires a nulling interferometer. 
Such an instrument  must be efficient in a large wavelength domain  in order to have the capability to detect simultaneously the infrared spectral features of several bio-tracers: CO$_{2}$, O$_{3}$ and H$_{2}$O. }
% Aims
{
A broad wavelength range can be effective, provided that an achromatic phase shift of $\pi$ can be implemented, with an accuracy good enough for achieving a deep nulling at all considered wavelengths.
 A new concept for designing such an achromatic phase shifter is presented here. The major interest of this solution is that it allows a   simple  design, with essentially one device per beam.
}
% Methods
{
 The heart of the system consists in two cellular mirrors where each cell has a thickness   which  introduces, {for a   given central wavelength,} a phase shift of $(2k + 1)\pi$ or of  $2k \pi$ on the fraction of the wave it reflects.
  Each mirror is put in one of the collimated beams of the interferometer.
 Because of the odd/even distribution, a destructive interference is produced on axis for the central wavelength when recombining the two beams.
 Now,  if the number of cells of a given thickness follows a rather simple law, based on the Pascal's triangle, then we show that the nulling is also efficient for  a wavelength not too far from the central wavelength.
}
% Results
{ The effect of achromatization is the more efficient the larger the  number of cells is.
 For instance, with two mirrors of 64$\times$64 cells, where the cells  phase shift  ranges  between -6$\pi$ and +6$\pi$, one reaches a  nulling of  $10^{-6}$ on a wavelength range $[0.6 \lambda_{0}, 1.25\lambda_{0}]$, i.e. on more than one complete octave.
 This is  why we claim that this device produces a quasi-achromatic phase shift : especially, it could satisfy the specifications of space mission as DARWIN. 
 In a second step, we study the optimum way to distribute the cells in the plane of the pupil.
 The most important criterion is the isolation of the planet image from the residual image of the star.   
 Several algorithms are presented, one being especially efficient and we present the nulling performances of those various configurations. 
 % last sentence suppressed 
 }{}
\titlerunning{The achromatic chessboard}
\authorrunning{D. Rouan, D. Pelat}

   \maketitle

\section{Introduction}
There is a huge contrast (10$^{6 -10}$) between a star and a planet that orbits it, so that  directly 
detecting the planet requires to  cancel as much as possible the stellar light. Today, it is generally admitted that there are two main instrumental paths to 
achieve this goal : one is coronagraphy on a single telescope \citep[e.g.][]{Guyon-a-07} -- generally in the visible to near-IR domain --, the other one is nulling interferometry -- proposed essentially in the thermal infrared --
using at least two telescopes recombined in a clever way  \citep[e.g.][]{Leger-a-96, Woolf-a-97}. This paper pertains to the second category. 
\cite{Bracewell-a-78} was the first to propose the concept of a nulling interferometer, where the light collected by two telescopes is coherently recombined, after  that a $\pi$ phase shift be applied on one of the  two arms of the interferometer.  
The system of fringes projected onto the sky shows then a central dark fringe : 
if the star image is put on this central  dark fringe, it is strongly attenuated (actually, the stellar photons are all sent in the second output of the interferometer). If, at the same time, the planet is on a bright fringe, i.e. a region of transmission unity, then it can be detected -- in principle -- in a much more efficient way, especially because of the reduced photon noise.

Obtaining an achromatic $\pi$ phase shift is one of the main keys of success because  {\it a)}  the wavelength domain is broad  where the set of a reputed unambiguous spectroscopic signature of life, namely  CO$_{2}$ (15 $\mu$m), O$_{3}$ (9.5 $\mu$m) and H$_{2}$O (5-8  $\mu$m),  is to be found  \citep{Leger-a-96,Ollivier-a-07}, {\it b)} in photons starving experiments such those considered here,  a design based on a monochromatic detection is out of question.  

Several solutions have been presented in order to approach an achromatic $\pi$ phase shift in a large domain of wavelength : for a review, see \cite{Rabbia-p-04}. The three main solutions are the stacking of dielectric plates of different indexes and dispersion \citep{Mieremet-p-00}, the crossing of a focus \citep{Rabbia-p-05} and the use of a periscope optical train to reverse the electric field \citep{Serabyn-a-01}. The first solution provides only a quasi-achromatization since it relies on an approximate linearization of the optical path {\it vs} wavelength. The two last methods provide in principle an intrinsic achromatic phase shift, but which is strictly limited to $\pi$; instrumentral effect (differential polarization for instance) may also affect the resulting null. The Two first methods have the characteristic of introducing a specific sub-system in one arm only of the interferometer, resulting in an assymetric design with the drawback of a difficult  balancing of the two arms.    Finally one notes that all those solutions require the use of several optical components, requiring in general to fold the beams several times.  Is there a way of reaching a $\pi$ phase shift with a {  simple optical component, while keeping the interferometer design symmetric.} We present in the following a new concept that has this ambition.  

 Recently \cite{Rouan-p-03a,Rouan-p-04a,Rouan-p-06, Rouan-a-07}  presented a method to define interferometric configurations of telescopes that provide a $\theta^{\rm{n}}$ nulling function for any even value of n. Achieving a rather large value of n  is  important in order to produce a trough shape of the nulling function and thus a good cancellation of the leaks due to the finite angular size of the star. 
The principle is based on a remarkable property discovered by \cite{Prouhet-a-1851} about
a peculiar partition in two sets of the N = $2^L$ first integers, done  according to the Prouhet-Thu\'e-Morse sequence.
{This property is such that, selecting half of the N telescopes according to this sequence and applying to the output of each one a $\pi$ phase shift, makes possible the cancellation of the (L-1) first terms of the developement in $\theta$ of the recombined amplitude, thus leading to the wished $\theta^{2L}$ behaviour of the intensity.}

Here, we remain in the same spirit while tackling the question of the achromatisation of the $\pi$ phase shift. The idea is to try to apply a similar  principle of cancelation of the first terms of a Taylor's development of the amplitude, {  but now with respect to the relative wavelength difference $\Delta \lambda / \lambda_{0}$.}
The central component of the new concept is a chessboard mirror, i.e. a mirror made with a pattern of cells of different thickness.  

In this first paper, we  consider only the case of an interferometer with two square pupils telescopes and we establish mainly the theoretical grounds on which this new concept lies. We show that there is indeed a peculiar distribution of the cells thickness on the chessboard mirror that fulfils our goals of a quasi-achromatization of the $\pi$ phase shift. We also show that {  an adequate distribution of the cells on the mirror's surface can add more nulling power. }  We depict several configurations, especially one, that have good properties with respect to the detection of a planet. We then study the performances that a perfect device would feature and show that it would satisfy the specification of the currently proposed missions, such as Darwin for instance \citep{Leger-a-07}.  Several appendixes {detail the formalism and the mathematical background  for the readers who will be keen on a deeper investigation.  A second paper will address the questions of the performances in a real world where a device cannot be perfect, and will present the first results of an experimental demonstrator. }

\section{The concept of the chessboard phase shifter}

\subsection{ Formulation of the problem}

\begin{figure}
\centering
\includegraphics[width=7.5cm]{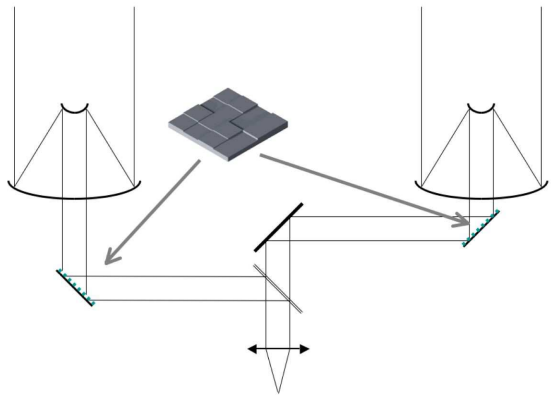}
\caption{The optical scheme of the nulling interferometer in coaxial configuration that we consider.  \label{fig:config}}
\end{figure}

\begin{figure}
\centering
\includegraphics[width=5cm]{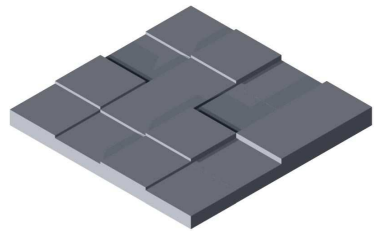}
\caption{An example of  cellular mirror we introduce in one arm of the interferometer. This is just to give an idea : the number of cells and their  arrangement are arbitrary.  \label{fig:miroir3D}}
\end{figure}

Let's consider an interferometer made with two identical afocal telescopes {  whose output beams are recombined in the so-called co-axial mode thanks to a beam-splitter.
The cartoon of Fig.\ref{fig:config} depicts such an optical scheme.}  To simplify the design and forthcoming computations, we assume that the shape of  each telescope's  pupils is {  limited by} a square.  In a collimated beam issued from the first telescope, we introduce a square cellular mirrors (alternatively transparent plate could be used). This cellular mirror has a {\it chessboard} design in the sense that it features n$\times$n square cells, arranged in column and rows as in a chessboard. The cells have  different thickness, however all {  are chosen to produce an optical path difference (o.p.d. hereafter) that is an odd multiple of $\lambda_{0}/2$}, where $\lambda_{0}$ is  the central wavelength. Fig. \ref{fig:miroir3D} gives an example of what a 4$\times$4 cellular mirror could look like.  
Symmetrically, a cellular mirror where cells {  produce  o.p.d. which are even multiples of $\lambda_{0}/2$} is put into the output beam of the second telescope.   {  By design}, each cell introduces then, on the fraction of the wave at wavelength $\lambda_{0}$ it reflects, a phase shift of $(2k + 1)\pi$ or of  $2k \pi$.   Hereafter, we call respectively {\it odd}  and  {\it even} the  two cellular mirrors. Of course the number of cells is the same on the two mirrors and their geometry is unique (square).
 
{  Note that a Fizeau (or multi-axial) recombination, i.e. done at the image plane level and not at the pupil level, could also be considered, with the strong advantage to keep the design fully symmetric. }It has been shown \citep[ e.g.]{Buisset-p-06, Wallner-p-04} that with this design, nulling  is as efficient as in a more classical {  co-axial} recombination (Michelson type), provided that  the star image (i.e. the  PSF)  is formed on a single mode fiber optics that will make the overlap integral of the product between the incident electric fields and the fundamental mode distribution of the fiber. In that case, the nulling comes from the fact that the amplitude is anti-symmetric with respect to the y axis.  However, in the following, we {  will stick to the co-axial configuration and especially will show images of the resulting PSF for that type of recombination. The effect of nulling on the putative planet detection is seen with a much better contrast, most of the stellar light intensity being cancelled. }

The {  amplitude integrated on the whole pupil area},  at the output of one {  chessboard} mirror  is : 

\begin{math}
a = \sum_{{k=1}}^{{N}} \exp(j  \phi_{{k}})
\end{math}

where  $\phi_{{k}}$ is the phase difference between one cell and a reference cell  :  { $\phi_{{k}} =  2 \pi  d_{{k}} / \lambda $, and  $d_{{k}}$ is the  o.p.d. produced by  cell k. 
 Let's now assume that  each  o.p.d.  is a  multiple of $\lambda_{0} / 2$}, where $\lambda_{0}$ is the  central wavelength. The phase  then reads : $\phi_{{k}} =  \pi  n_{{k}}\lambda_{0} / \lambda $.  If $\lambda$ is within a rather narrow range with respect to $\lambda_{0}$, then one can write : $\phi_{{k}} =  \pi  n_{{k}} (1 -  \Delta \lambda / \lambda_{0} )$.

 If now, on one cellular mirror all the {  o.p.d. are even multiple of $\lambda_{0} / 2$,  and on the other mirror, they are odd multiple of   $\lambda_{0} / 2$, then the complex amplitudes take} respectively the form :\\

  \begin{math}
a_{+}  = \sum_{{k=1}}^{{N}}  \exp(j  \pi  m_{{k}}  \Delta \lambda / \lambda_{0} ))
\end{math},  where $m_{{k}}$ are even integers and

 \begin{math}
a_{-}  = \sum_{{k=1}}^{{N}} - \exp(j  \pi  n_{{k}}  \Delta \lambda / \lambda_{0} ))
\end{math},  where $n_{{k}}$ are odd integers.

 {   When, after reflection on the cellular mirrors, the two beams  are recombined, then the integrated amplitude $a$ of the  wave at the output of the recombiner (or, this is equivalent, at the focus of the final lens) is  :}

  \begin{math}
a = a_{+}  + a_{-}  =  \sum_{{k=0}}^{{N}}  \exp(j  \pi  m_{{k}}  \Delta \lambda / \lambda_{0} )) - \sum_{{k=0}}^{{N}}  \exp(j  \pi  n_{{k}}  \Delta \lambda / \lambda_{0} ))
\end{math}

{  Let's note  $\phi =   \pi  \Delta \lambda / \lambda_{0} $ and let's develop } each exponential in series of $(n_{{k}}\phi)^n$, so that the two terms of the above expression become :  

\begin{math}
a_{+} = \sum_{{k} } 1 +  j\phi\times\sum_{{k} } m_{{k}}  -  
\phi^{2}\times\sum_{{k}} m_{{k}}^2 -
 j\phi^{3}\times\sum_{{k} } m_{{k}}^3 + ...  
\end{math}

and

\begin{math}
a_{-} = - \sum_{{k} } 1 -  j\phi\times\sum_{{k} } n_{{k}}  +  
\phi^{2}\times\sum_{{k}} n_{{k}}^2 +
 j\phi^{3}\times\sum_{{k} } n_{{k}}^3 + ...  
\end{math}

Clearly, since the number $N $  of cells is the same for the two mirrors, the first term of each expression cancels out mutually. In order to cancel out the other terms, at least up to a given order $O$, we need to find one set $\{m_{{k}} \}$ of even integers and one set  $\{ n_{{k}} \}$ of odd integers, such that $\sum_{{k}} m_{{k}}^{o} = \sum_{{k}} n_{{k}}^{o}$ for all $o \leq  O$.  Note that several of the  $ m_{{k}}$ (respectively of the $n_{{k}}$) can be identical.  This problem is known as the Prouhet-Tarry-Escott problem \citep{Borwein-a-94}, with here the additional peculiar condition on  the odd/even repartition of the integers. 

The good news is that solutions do exist for any value of  $O$ ! Even more, there are several solutions for  a given value of $O$ and the number of solutions increases very rapidly with $O$. For instance, let's consider a nulling up to $O=2$ : we are looking for at least one set of even integers  $m_{{k}}$ and one set of odd integers $n_{{k}}$, such that : \\
$\sum_{{k} } m_{{k}}  =  \sum_{{k} } n_{{k}}$   and 
$\sum_{{k}} m_{{k}}^2 = \sum_{{k}} n_{{k}}^2  $\\
One solution (that can be found for instance by trying all possible combinations of first odd and even integers) is :\\
$\{m_{{k}}\} =  \{2, 2, 2, 4\}$ , 
$\{n_{{k}}\} =  \{1,3,3,3 \}$\\
One checks that  2 + 2 + 2 + 4 = 1 + 3 + 3 + 3 = 10 and that 
$2^{2} + 2^{2} + 2^{2} + 4^{2} = 1^{2} + 3^{2} + 3^{2} + 3^{2} = 28$

At this point, one remarks that an infinity of solutions can be built from a given one : it suffices  to add any integer to all elements of the two sets. For instance the sets   $\{m_{{k}}\} =  \{4, 4, 4, 6\}$ , 
$\{n_{{k}}\} =  \{3,5,5,5 \}$, obtained by adding 2 to each element of the solution given as first example, constitute a valid solution. This is so because  \\
 $\sum_{{k}} (m_{{k}}+ L)^{o} = \sum_{{k}} m_{{k}}^{o} + o L m_{{k}}^{o-1} +   {o\choose 2}  L^2 m_{{k}}^{o-2} + ... + L^o$ , where any   elementary {\it even sum} 
 $\sum_{{k}}  {o\choose i}  L^i m_{{k}}^{o-i}$
 is equal to the equivalent {\it odd sum} :
 $\sum_{{k}}  {o\choose i}  L^i n_{{k}}^{o-i} $

From the optical point of view, it's clear that all those solutions are totally equivalent : this is nothing but  just adding everywhere a same phase shift (or piston). In the following, we will use this property of invariance by translation.   Hereafter we'll consider only solutions where at least one cell has a null thickness (0) : this means  simply that it corresponds  to the reference level of the mirror surface. 

In the following section, we examine how proper solutions can be built in a coherent and systematic way for any value of $O$.

\subsection{ The various solutions}

As we said, there is a rapidly increasing number of solutions when $O$ increases. For instance, for $O=3$, a systematic search, gives 16 solutions when the number of cells per mirror is set to 32.  However, all those solutions are not equivalent from the instrumental point of view.
We are looking for solutions which are best adapted to our optical problem. One natural condition is that the dispersion of the thickness on the surface of the mirror is not too large : obviously if the cellular mirror  comprises a few cells with thickness much larger than the other, then there will be both a problem of manufacturing of the mirror (steep edges) and of mutual shadowing by the cells.  For instance, the solution  $\{m_{{k}}\} =  \{4, 6,  6,  0, 0, 0, 0, 0\}$ , 
$\{n_{{k}}\} =  \{1,1,1,1,1,1,1, 9 \}$\\  is valid for $O=2$, but the odd mirror exhibits a {\it tower} of {  high thickness  emerging from a flat {\it pond} and this is clearly not desirable. }
 
   \begin{figure}
\centering
\includegraphics[width=8cm]{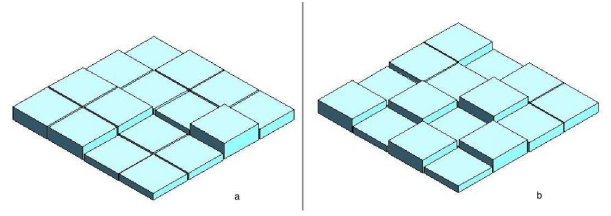}
\caption{ { Example of a pattern of cells that is valid for an on-axis nulling but that will clearly have drawbacks. The  X-Y distribution a) will produces aberrations (e.g. tilt) while the b) one is more homogeneous and will be preferred.}  \label{fig:confs-bad}}
\end{figure}

To avoid large dispersion in thickness, one can put as a condition that the ordered set of thickness, increases {  by one unit from one cell to the other}. In other terms we want to use all thickness in a set of incrementing integers.   This is a guarantee of a rather low variance of the thickness on the surface of the mirror.  

\subsection{ The optimized solution}

We found an elegant and singular solution which  satisfies this \textit{maximum flatness} condition and which  is based on the Pascal's triangle (the rule to derive recursively the binomial coefficients). More precisely, for a nulling up to $\phi^O$, we consider mirrors with $2^{O}$ cells, the number of cells of thickness $k$ being the binomial coefficient  $O \choose 2$. A clear way to present the solution is simply to write the first integers fro 0 to $O$ on one line and the binomial coefficients of order $O$ on the next line.
\begin{center}
\begin{tabular}{|c||c|c|c|c|c|c|c|}
o.p.d. introduced by cell & 0 & 1  &   2               &   3             &  ...&  $O-1$ &   $O$ \\
Number of cells   &1  & $ O$ & $O \choose{2}$ &$O \choose{3}$ & ... & $O$     & 1\\
\end{tabular}
\end{center}
{  The first line gives the o.p.d. introduced by the cell (in unit of the elementary step $\lambda_{0}/2$) and the second line gives the number of cells of this kind. }
For instance, if $O=6$ : 
\begin{center}
\begin{tabular}{|c||c|c|c|c|c|c|c|}
Thickness of cell & 0 & 1 &   2&   3&   4&   5&   6 \\
Number of cells &1 &  6 & 15 & 20 & 15 & 6 & 1\\
\end{tabular}
\end{center}

 This means that the even  mirror is built as follows : a) it is made of $O/2 +1$ (if $O$ is even)  or $(O+1)/2$ (if $O$ is odd) subsets of cells;  b) all cells in the subset $k$ {have an o.p.d. equal to $k$}; c)  the number of cells in the  subset $k$ is given by the coefficient of the binomial, $O\choose{k}$.  All even values of $k$, such that $ 0 \leq k \leq O$  are thus found at the surface of the even mirror.  For instance, 
1 cell of null o.p.d. ,  $ {O \choose 2} = O(O-1)/2$ cells of o.p.d.  2, $ O\choose{4}$ cells of o.p.d.  4, ..., 
1 cell of o.p.d.  $O$ if $O$ is even, or $O$ cells of o.p.d.  $O-1$ if $O$ is odd.  Symmetrically, the  
odd mirror is built with the same rule, but with $k$ being odd;  it exhibits thus :  
O cell of o.p.d.  1,  $ O \choose{3}$ cells of o.p.d.  3, $ O\choose{5}$ cells of o.p.d.  5, ..., 
1 cell of o.p.d.  $O$ if $O$ is odd or $O$ cells of o.p.d.  $O-1$ if $O$ is even. Note that there are $O/2$ (if $O$ is even)  or $(O+1)/2 $ (if $O$ is odd) subsets of odd cells.

One can demonstrate that this solution do satisfy the condition of nulling :  $\sum_{{k}} m_{{k}}^{o} = \sum_{{k}} n_{{k}}^{o}$ for all $o \leq O$ (see Appendix).

\section{ X-Y distribution of cells in the pupil : which criterion ?}

If the optimum solution provides the distribution in $z$ of the cells, it does not say anything on the distribution in $x$ and $y$, i.e. on the surface of the mirrors. Once again, some common sense can be exercised to select one with respect to another. For instance, if all the large thickness are on one side of a mirror,  or if there is a continuous gradient of thickness across the surface, then this is equivalent to introducing some tilt  on the wavefront, a bias that is not desirable (see Fig.\ref{fig:confs-bad}-right for an example of configurations with a same z-distribution but different x-y distributions, one being better than the other). A first condition that can be stated is that the global tilt be zero. This can be expressed by requiring that the first moments of the phase of the reflected wavefront is zero :\\
 $\sum_{{i,j} } {i} \times  m_{{i,j}} = 0$ and $\sum_{{i,j} } {j}  \times  m_{{i,j}} = 0$, \\ 
 where i, j are the coordinates of a cell in a chessboard pattern, the origin being at the center of the mirror.  
 This does not seem to be a constraint hard to fulfill. To go further, one can ask that higher moments of the phase be equal to zero, so that higher order aberrations are cancelled  : \\
  $\sum_{{i,j} } {i}^o \times  m_{{i,j}} = 0$ and $\sum_{{i,j} } {j}^o  \times  m_{{i,j}} = 0$ \\
  In a final step, all moments, including the cross terms could also be considered : \\
  $\sum_{{i,j} } {i}^o \times   {j}^q  \times  m_{{i,j}} = 0$ \\
  
  Another criterion is to consider the image plane, i.e. the plane conjugated to the pupil plane through the Fourier transform (in an on-axis recombination as above-mentioned).   We can look for patterns of cells, such that the  nulling for the starlight extends as far as possible from the centre, for instance along the two main axis, with of course the less of attenuation  for the planet.   
  
  Let's assume that the cells on the even and odd mirrors  are distributed according to repectively a matrix P and a matrix Q, where each element e$_{i,j}$ is the integer determining the o.p.d.  of the cell ($i,j$). 
  
  At a wavelength  $\lambda \ne \lambda_{o}$, the Fourier transform ( hereafter noted $~^{t}\{ \}$) of the combined pupil reads : \\
    \begin{math}
  ^{t}\{ a_{+} + a_{+}   \} =  ^{t}\{(\exp j  P \phi -   \exp j Q \phi) \star cell  \}
  \end{math}, \\
  where $cell$ means a function  whose value is 1 inside a square of the size of  one cell and 0 outside. Note that with this condensed notation, $\exp j P \phi$ means in reality the  2D Dirac's comb with a period of one cell width and an amplitude  $\exp j P_{k,l} \phi$.  Using the properties of the convolution $\star$, this expression becomes :
  
    \begin{math}
  ^{t}\{ a_{+} + a_{+}   \} =  ^{t}\{\exp j P\phi -   \exp j Q\phi \} \times PSFC   \end{math}, \\ 
  where $PSFC$ is the point spread function given by a unique cell. Since the term $PSFC$ corresponds just to the envelope that modulates the resulting PSF, we can restrain the study to $^{t}\{ \exp j P\phi -   \exp j Q\phi \}$, the term which carries the information on the distribution of the cells.
Because we are dealing with Dirac's combs, the Fourier transforms reduces to a discrete sum :
  
 $^{t}\{\exp j P\phi -   \exp j Q\phi \} =   \sum_{k,l} ~( \exp j p_{k,l} - \exp j q_{k,l} )~ \exp j \alpha (k x + l y) $, \\   
    where $p_{k,l}$ and $ q_{k,l}$ are the elements  of matrix $P$ and $Q$ respectively, and $\alpha$ is a coefficient resulting from the Fourier transform. After a Taylor's development, this 
expression  can be written :
  
    \begin{math}
  \sum_{k,l} ~   [ j  \phi  (p_{k,l} - q_{k,l}) -  \phi^2 ( p_{k,l}^2 - q_{k,l}^2) -j \phi^3 ( p_{k,l}^3 - q_{k,l}^3) + ... ] ~\exp j \alpha (k x + l y)
  \end{math}. \\
   Considering this expression, we see that a sufficient condition to get a null contribution at any point of the $y$ axis $(x = 0)$,  is to have simultaneously : 
   
$ \sum_{k} (p_{k,l} - q_{k,l} ) = 0, ~ \sum_{k} (p_{k,l}^2 - q_{k,l}^2) = 0,  ~\sum_{k} (p_{k,l}^3 - q_{k,l}^3 ) = 0, $ ~ etc.  for any value of $l$.  \\
We remark  that this is the same condition, applied here to each column, that we found for the whole pupil, on the power of sets of odd and even numbers, i.e. $\sum_{k} m_{k}^{o} = \sum_{k} n_{k}^{o}$ for all $o \leq O$.

 This means that if we can distribute the cells along the columns of the odd and even mirror, so that for each column the condition      
 $\sum_{k} m_{k}^{o} = \sum_{k} n_{k}^{o}$ is satisfied up to an index $O$, then the amplitude is not only very low at the centre, but also  along the $y$ axis. Of course, the same result applies  along the $x$ axis 
 if the condition can be satisfied for all lines. This does not guarantee that outside those two main axis the nulling will be as deep, but by continuity of the PSF, there  is a good chance that it remains fairly good. We will see below that we found indeed such distributions of cells that satisfy this condition, both along the lines and the columns {  (see also Appendix D for a different but complementary approach)}. 

\subsection{ The optimized solutions}

We found several ways to geometrically  distribute the cells on the mirror's surface, following one or the other of the prescriptions established in the previous section.  We describe hereafter three of them, each with its advantages and limits. The first one corresponds to the first criterion specifying that the first moments in the pupil  must be zero, while the two other patterns satisfy  the second criterion on the  nulling along the $x$ and $y$ axis in the image plane. Other solutions, which are essentially variants of those three have been also explored and the section on performances will include them. 

\subsubsection{First method}

 This first method tries to fulfil the condition on minimizing the moments on the wavefront after the reflection on the chessboard mirror; it  is based on a recurrence scheme to build the pair of mirrors of order $O$ from the pair of order $O-1$. 
   Let's assume that three square  matrix $B$, $P$, $Q$ are defined with dimension $2^{O-1} \times 2^{O-1}$. The even matrix is $P$ and the odd one is $Q$, while  $B$ is an auxiliary matrix.    We then build the matrix of dimension $2^{O} \times 2^{O}$ as follows :
   
\begin{center}
\begin{tabular}{|c||cc|}
\hline
$ P _{O}$ & $P _{O-1}$  & $Q_{O-1}  - B_{O-1}$  \\
 &  $-Q_{O-1} +B_{O-1}$ &  $-P _{O-1}$ \\
 \hline
 $Q_{O}$ & $Q_{O-1}$ & $P _{O-1} - B_{O-1}$ \\
   & $-P _{O-1} + B_{O-1}$ & $-Q_{O-1}$\\
   \hline
  $B_{O}$ & $B_{O-1}$ & $-B_{O-1}$ \\
    & $-B_{O-1}$ & $B_{O-1}$ \\ 
\hline
\end{tabular}
\end{center}

As a starting point we choose : $P $=0, $Q$=1, $B$=1. So that at the first step :

\begin{center}
\begin{tabular}{|c||cc|}
\hline
$ P _{1}$ & 0   & 0    \\
 &  0 & 0  \\
 \hline
 $Q_{1}$ & 1 &  $-1$ \\
   &   1 & $-1$ \\
   \hline
  $B_{1}$ & 1 & $-$1 \\
    & $-1$ & 1 \\ 
\hline
\end{tabular}
\end{center}

\begin{figure}
\centering
\includegraphics[width=6cm]{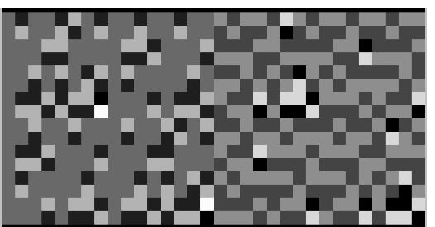}
\caption{  Example of pattern of cells in the case of the first method, with $O$ = 4. The grey level codes the o.p.d.  of the cells. \label{fig:DidierB}}
\end{figure}

 Fig. \ref{fig:DidierB} shows an example of the pattern of cells obtained with that method and Fig.\ref{fig:ComparaisonPSF}-a displays the resulting PSF with a planet 10$^{-6}$ fainter than the star. 
 
One can verify that the first moments of the wavefront, up to the order $O$, do satisfy the condition of being null. 
%Note however,  that this is not true for the cross-terms : we did not find a solution satisfying this additional condition.

\subsubsection{Second method}   

The aim of this distribution is to produce an efficient nulling along both the x- and y-axis in the image plane. 
 
Here, we start from the remark that if one writes the first $2^O$ integers in their binary form, and associate to each integer the number of bits set to 1, the frequency of those numbers  are distributed according to the binomial coefficients. For instance   if $O=3$ : 
\begin{center}
\begin{tabular}{|c|c|c|c|c|c|c|c|c|}
Integer & 0 & 1 &   2&   3&   4&   5&   6 & 7  \\
Binary &000& 001&010&011&100&101&110&111\\
\# of 1 & 0 & 1 & 1 & 2 & 1 & 2 & 2 & 3 
\end{tabular}
\end{center}
The occurence of bits set to 1  is indeed ${3\choose 1} = {3\choose 2} = 3$ for one or two bits set,  and ${3\choose 0} = {3\choose 3} = 1$  for zero or three bits set.

If we replicate the elements of this line through all permutations,  we will keep the property of the Pascal's triangle. We can then build easily a square matrix with  $2^O$ lines of this kind. However, we wish to do it so that we also produce columns that share the same property.   
 One practical way to obtain this result  is to write $i \oplus j$  (exclusive OR) at the intersection of line $i$ and column $j$. The number of bits set to 1 on the resulting number gives then the o.p.d.  of the cell in steps of $\lambda_{o}/4$. An example is given below : 

\begin{center}
\begin{tabular}{|c|c|c|c|c|c|c|c|c|}

  &000 &  001 & 010 & 011 & 100 & 101 & 110 & 111 \\
\hline
000 & 0 & 1 & 1 & 2 & 1 & 2 & 2 & 3 \\
001 & 1 & 0 & 2 & 1 &2 & 1 &3 & 2 \\
010 & 1 & 2 & 0 & 1 & 2 & 3 & 1 & 2 \\
011 & 2 & 1 & 1 & 0 & 3 & 2 & 2 & 1 \\
100 & 1 & 2 & 2 & 3 & 0 & 1 & 1 & 2 \\
 etc . &&&&&&&&\\
\end{tabular}
\end{center}
 
 One can check that : {\it a)} each line or each column contains the right number of cells following the Pascal's triangle rule; {\it b)} the lines and columns are distributed exactly the same way (the matrix is symmetric with respect to the second diagonal).
  After this operation is done,  the odd and even numbers are interleaved.  The odd and even mirrors are then built by simply adding 1 to respectively the even cells or  the odd cells, as illustrated in the following table :
  
  \begin{center}
\begin{tabular}{|c|c|c|c|c|c|c|c||c|c|c|c|c|c|c|c|}
\hline
\multicolumn{8}{|c||}{even mirror} & \multicolumn{8}{c|}{odd mirror} \\
\hline
 0 & 2 & 2 & 2 & 2 & 2 & 2 & 4 & 1 & 1 & 1 & 3 & 1 & 3 & 3 & 3\\
 2 & 0 & 2 & 2 &2 & 2 &4 & 2 & 1 & 1 & 3 & 1 &3 & 1 &3 & 3 \\
 2 & 2 & 0 & 2 & 2 & 4 & 2 & 2 & 1 & 3 & 1 & 1 & 3 & 3 & 1 & 3 \\
 2 & 2 & 2 & 0 & 4 & 2 & 2 & 2 & 3 & 1 & 1 & 1 & 3 & 3 & 3 & 1\\
 2 & 2 & 2 & 4 & 0 & 2 & 2 & 2  & 1 & 3 & 3 & 3 & 1 & 1 & 1 & 3 \\
...&...&...&&&&&\\
 \hline
 
\end{tabular}
\end{center}
  
   We know, thanks to the property of invariance by translation, that the condition on the sum of powers is kept. Because of that and because of property {\it a)} and  {\it b)}, the searched condition on the sum of powers along each column or each line is thus satisfied.
 
\begin{figure}
\centering
\includegraphics[width=6 cm]{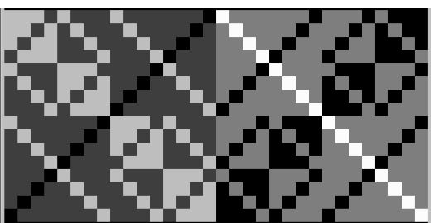}
\caption{  Example of pattern of cells in the case of the first method, with $O$ = 4. The grey level codes the o.p.d.  of the cells. \label{fig:binaryswap}}
\end{figure}

 Fig. \ref{fig:binaryswap} shows an example of the pattern of cells obtained with that method and Fig.\ref{fig:ComparaisonPSF}-b displays the resulting PSF with a planet 10$^{-6}$ fainter than the star. 
 
  \subsubsection{Third method}
 The third method  aims also at  producing an efficient nulling along both the x- and y-axis, but is  now based on a recurrence scheme which is applied to a single matrix $R_{O}$ (with  $R_{0} = 0$), as follows :
 
\begin{center}
\begin{tabular}{|c||cc|}
\hline
$ R _{O}$ & $R _{O-1}$  & $R_{O-1}  + (-1)^{O}$  \\
 &  $R_{O-1} - (-1)^{O}$ &  $R _{O-1}$ \\
\hline
\end{tabular}
\end{center}
For instance : 
\begin{center}
\begin{tabular}{|c||cccc|}
\hline
$ R _{1}$ & 0   & -1 &  &  \\
 &  1 & 0  &&\\
 \hline
$ R _{2}$ & 0   & -1  & 1 & 0 \\
 &  1 & 0  & 2 & 0 \\
 & $-1$ & $-2$ & 0 & $-1$ \\
 & 0 & $-1$ & 1 & 0 \\
\hline
\end{tabular}
\end{center}

\begin{figure}
\centering
\includegraphics[width=6cm]{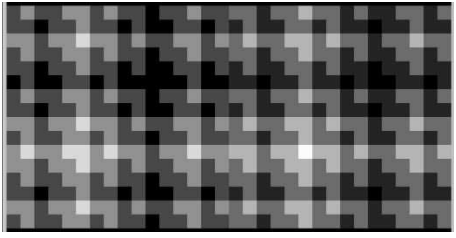}
\caption{  Example of pattern of cells in the case of the third method, with $O$ = 6. The grey level codes the o.p.d.  of the cells. \label{fig:DanielC}}
\end{figure}

To extract respectively  the odd and even matrix, one simply add 1 to the even elements or to the odd elements in matrix $R$, as in method 2.  One can check that, again, the searched condition on the sum of powers along each column or each line is satisfied.
 Fig. \ref{fig:DanielC} shows an example of the pattern of cells obtained with that method and Fig.\ref{fig:ComparaisonPSF}-c displays the resulting PSF, assuming a planet 10$^{-6}$ fainter than the star. 
 
 In fact, two versions of this pattern exists : one which is anti-symmetric and produces an assymmetric PSF and one which is symmetric and produces symmetric PSF. It seems, for practical reasons (e.g. alignment of the optical setup) that the second pattern would be preferable.  

\section{Simulation and predicted performances}

The most important question is now to evaluate if the achromatic phase shifter we propose is indeed efficient, both in terms of nulling and of wavelength range. For that purpose
We have developed a numerical simulation to assess the respective performances of the various configurations we have found.
Essentially, it consists in : a) building a pair of phase masks for a given type and  a choice of dimensions ;
b) assuming a source exactly on axis, to perform a FFT on the sum of the complex amplitudes on the two masks,  for a given shift of wavelength with respect to the central one : this produces the resulting nulled image of the star, of course after taking the square of the modulus ;
c) to add a phase shift, generally $\pi$, on one mask,  in order to mimic the presence of a planet 
that would be on  a constructive interference fringe and again to perform the  FFT and take the square of the modulus.

\begin{figure}
\centering
\includegraphics[width=4cm]{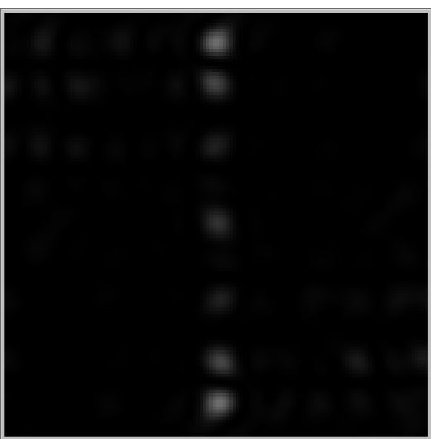} \hfill
\includegraphics[width=4cm]{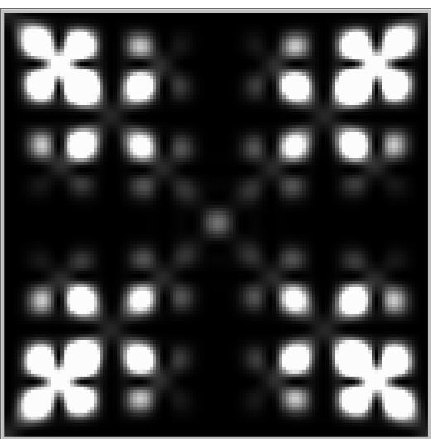}  \hfill
\includegraphics[width=4cm]{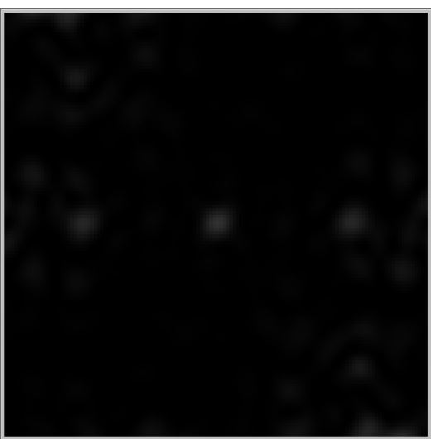}
\caption{  Comparaison of the planet image obtained with three different methods for generating the chessboard mirror pattern.  A log scale is used. The planet, which has a flux $10^{-6}$ times fainter than the  star, is the pointlike source at centre; it is surrounded by residuals of the starlight.  The same cuts have been used, so that the difference in intensity seen on the planet is meaningful.  \label{fig:ComparaisonPSF}}
\end{figure}

In Fig. \ref{fig:ComparaisonPSF} we compare the different images of the planet -- assumed to be at a level of $10^{-6}$ of the stellar flux -- and residuals of the star,  obtained fot the typical case $( \lambda - \lambda_{o}) / \lambda_{o} = 0.2$.

The assessment  is then done : a) on the quality of the nulling versus the wavelength, for instance in computing the residual  flux in a circular diaphragm centred on the star position; b) on the magnitude  of the unavoidable partial extinction of the planet; c) on the wavelength range where a good nulling is obtained. 

The simple metric we have used to  compare the different configurations is to compute the ratio of the residuals of the planet flux to the star flux, each being measured within one resolution element of a single telescope, i.e. an area of $\lambda /D$ around the main axis in the image plane. 
Another possible metric would be to estimate a signal to noise ratio  on the planet detection, assuming either  photon noise (planet + stellar residuals) or a noise proportional to the stellar residuals since it seems clear today that a lack of stability in the nulling will probably be the main source of noise in nulling interferometers  \cite{Lay-a-04}). 
 
 \begin{figure}
\centering
\includegraphics[width=9cm]{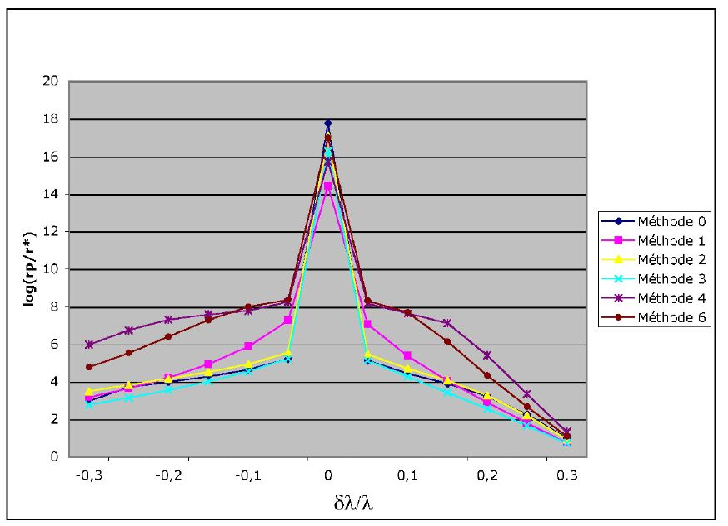}
\includegraphics[width=9cm]{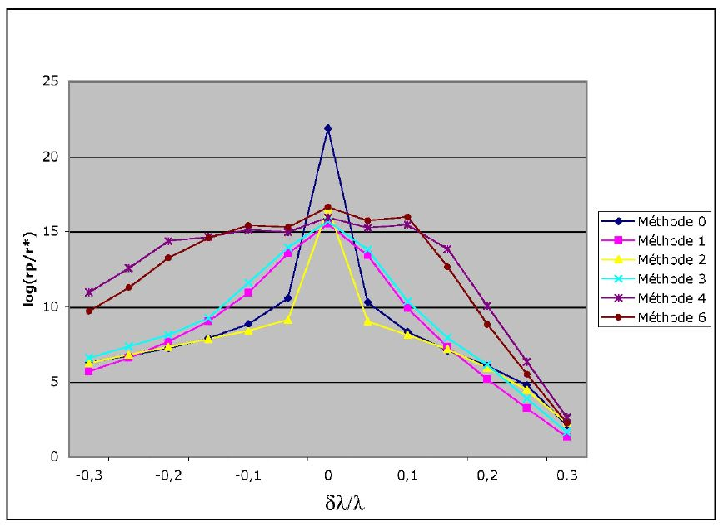}
\caption{Plot of the theoretical gain in planet to star contrast (log scale), as a function of the relative wavelength difference   $( \lambda - \lambda_{o}) / \lambda_{o}$ for different x-y configurations of the cells.  Top : mirrors with 8$\times$8 cells; bottom :  mirrors with 64$\times$64 cells.  The singular peak at the central wavelength  corresponds to a perfect nulling (gain $\infty$)  so that the point translate only the numerical accuracy. The different curves correspond to the different patterns of cells. The two curves noted 5 and 6 correspond to the two versions of the third method (see text).  Curve 3 and 4 corresponds to two variants of method 2 and  curves 1 and 2 to variants of method 1.    \label{fig:result_model}}
\end{figure}

We present in Fig. \ref{fig:result_model} the results of the model in terms of nulling efficiency vs $( \lambda - \lambda_{o}) / \lambda_{o}$ for  the different configurations we have tested and for two numbers of cells in the phase mask.  The following remarks can be done : 

First, it is specially rewarding to see that the specifications on nulling efficiency that a DARWIN type mission  would require (a contrast of 10$^{-6}$ and in fact a  specification of 10$^{-5}$ on the nulling efficiency), could be easily met with the solution we propose : e.g.  theoretical contrasts as high as 10$^{15}$ are predicted on  a significant wavelength domain. 

Second, the performance in terms of achromaticity is fair, since the level of 10$^{-6}$ is reached on the domain  [0.6 $ \lambda_{o}$ -- 1.25  $\lambda_{o}$] with the best configuration : this corresponds to more than one complete octave in wavelength.  It means that with only two devices it is possible to cover  the whole wavelength range of DARWIN.  We note also that there is an asymmetry of the useful wavelength domain : the performances are within the specifications in a broader domain on the short wavelength side.  

Third, the number of cells is important as shown by comparing curves at top (8$\times$8) and at bottom (64$\times$64)  in Fig.\ref{fig:result_model}. However, increasing this number beyond  64$\times$64 does not bring a very significant gain. 

 Fourth, among the different patterns of chessboard,  clearly the third method presents much better results -- at least when we consider the raw numbers -- since, for instance, it gives a contrast 10$^6$ times better than the next best configuration at   $(\lambda - \lambda_{o}) / \lambda_{o }= - 0.2$ . This does not mean that practically it would always be the best and a more thorough analysis is needed, taking into account deviation from a perfect system, once the device is manufactured.  
 
\section{conclusion}

We have presented  in this paper a new concept of a quasi-achromatic phase shifter, a functionnality which is at the heart of any nulling interferometer. We think that it can simplify the design and the setup of the interferometer since its relies on a {  simple component of bulk optics symmetrically introduced in the two arms of the interferometer.} In that sense, it 
could overcome certain limitations of other solutions  which use multi-components systems and/or {  complex} asymmetric design. 
The heart of the concept is a chessboard mirror with a large number of cells (typically 64$\times$64), put in the pupil plane. It is the peculiar distribution of the different cell thickness which is at the origin of the quasi-achromatic capability : it allows to cancel out the first terms of the Taylor's development in $\Delta \lambda / \lambda$ of the amplitude, up to any given order. Among the various distributions that satisfy this condition, a very unique one is especially optimum.
We also show that there is a peculiar x-y distribution of the cells which gives very good performances in terms of planet to star contrast.  
Reaching a very deep nulling on a broad wavelength range  is a mandatory condition  for detecting terrestrial planets with nulling interferometers and ultimately characterize them, so that this solution can be of interest only if it reach the proper performance of nulling in a broad enough domain and, indeed, we show that the specification of 10$^{-6}$ in contrast that a space mission as DARWIN is aiming at,  can be  reached  on a complete octave in wavelength in the best case of a 64$\times$64 chessboard mirror. 
 
Finally, we want to introduce a new terminology. The set of equations on which this new concept is based is of diophantine type (i.e. polynomial equations where solutions are integers), as was a similar set of equations in the question of very deep nulling interferometers that one of us  is exploring \citep{Rouan-a-07} since a few years;  as a consequence, we propose to call {\it diophantine optics} this new type of optics where peculiar relations between powers of integers are used to obtain very specific effects. 

\bibliography{HRA,HRA-a-DRouan,HRA-p-DRouan}
\bibliographystyle{aa}

\appendix

\section{The z-distribution based on Pascal's triangle}
  
We want to demonstrate that the condition   $\sum_{k} m_{k}^{o} = \sum_{k} n_{k}^{o}$ for all $o \leq O$  is fullfilled with the solution given by the Pascal's triangle.
We use a recurrence reasoning.
The nulling relation we want to demonstrate reads :\\
  \begin{math}
 \sum_{k=0}^{O} (-1)^k k^o  {O\choose k} = 0 
   \end{math}  for all $o \leq O$.\\
The expression on the left can be rewritten:\\
   \begin{math}
 \sum_{k = 0}^{O} (-1)^k k^{o-1}  k {O\choose k} 
   \end{math}  \\
  We use now the relationship $   k {O\choose k}  =  O {O-1 \choose k-1} $ to transform this expression in : \\
$ \sum_{k=0}^{O} (-1)^k k^{o-1}  O {O-1\choose k-1} $\\
 We note that when $o \ne 0$   the summation can start at $k = 1$, so that the equation to verify is : \\
  \begin{equation}
   \label{Eq:A1} 
    \sum_{k=1}^{O} (-1)^k k^{o-1}  O {O-1\choose k-1}  = 0   
  \end{equation} 
Now let's assume that the nulling relation holds for $p-1$ and $O-1$ \\
 $ \sum_{k=0}^{O-1} (-1)^k k^{o-1}   {O-1\choose k} = 0   $\\
 Let's use the property demonstrated in section 1 that any translation of the o.p.d.  keeps the relation true, so that, with a translation of $+1$: \\
 $ \sum_{k=0}^{O-1} (-1)^k (k+1)^{o-1}   {O-1\choose k}  =  0$ \\
 This can be rewritten :\\
 $\sum_{k=1}^{O} (-1)^k k^{o-1}   {O-1\choose k-1}  =0  $\\ 
 This is the same expression as in Equ.\ref{Eq:A1}, where it is just multiplied by $O$.
 This demonstrates that  if  the relation holds for $p-1$, then it holds for $p$. 
It's then sufficient to find a value of $p$ for which the equation is true, to prove that it is true for any $p$.
For $p=0$, the condition becomes :
   \begin{math}
 \sum_{k=0}^{O} (-1)^k  {O\choose k} = 0    
  \end{math}, 
 where the left part is actually the development of $(1 - 1)^O$  and is obviously 0.

\section{Image in the focal plane} 
In order to assess the device's performance after insertion of the phase grids into the beam, it is convenient to calculate analytically the image at the telescope focus.
This image is obtained, at wavelength $\lambda$, by squaring the Fourier transform module of the pupils.
One obtains formally : $\text{image}(u,v)=\text{module}^2(u,v)$.
We have :
\begin{equation}
\label{eqa1}
\text{module}(u,v)=\sinc(\tfrac{\pi}{n_0}ud)\sinc(\tfrac{\pi}{n_0}vd)\bigl|\Sum(\text{expUV}\cdot\text{expPQ})\bigr|\,.
\end{equation}
In this expression, $u$ and $v$ are the coordinates on the focal plane, $d$ is the size of the mirror, $n_0$ is the size of the phase grids and the definition \(\sinc(x) = {\sin x}/{x}\) was adopted.
The remaining term involves the two square matrices $\text{expUV}$ and $\text{expPQ}$.
The operation $\cdot$ between the two matrices is the elements by elements product \emph{not} the matrices product.
The term $\Sum(A)$ is just the sum of all the elements in a matrix $A$ and, as usual, $|z|=\sqrt{z\bar{z}}$, $\bar{z}$ being the complex conjugate.
The two above matrices in \eqref{eqa1} have format $n_0\times n_0$ and are given by :
\begin{gather}
\text{expUV} = e^{j\pi ud\delta/n_0}\otimes e^{j\pi vd\delta/n_0}\,,\\
\text{expPQ} = e^{j\pi P\Delta}e^{-j\pi uD} + e^{j\pi Q\Delta}e^{+j\pi uD}\,.
\end{gather}
We have adopted the convention that the exponential of a matrix is the matrix of the exponentials.
Now, $D$ is the distance between the two telescopes and $\delta$ is a vector of integers varying from $-(n_0-1)$ to $+(n_0-1)$ by step of 2.
For example if $n_0=2$ we have $\delta=[-3,-1,+1,+3]$.
The symbol $\otimes$ is the tensor product, accordingly $\text{expUV}$ is a square matrix constructed as a tensor product of two vectors (sometimes called the outer product or the cartesian product).
Finally the chromatic term $\Delta$ is:
\begin{equation}
\Delta = \frac{1}{1+\frac{\Delta\lambda}{\lambda_0}}\,,\quad \Delta\lambda=\lambda-\lambda_0\,.
\end{equation}

The above expression is valid for the star of the system.
For the planet, one must correct $\text{image}(u,v)$ with the relative monochromatic luminosity at wavelength $\lambda$ and, if one neglect the planet tilt on the telescope pupils, add the phase shift on one grid.
Typically, we set the $Q$ grid to $Q+1$, in order to obtain maximum planet detectivity through a $\pi$ phase shift.

\section{First order: the Prouhet-Pascal solution}
At first order approximation, we set $D=0$ and $\delta=0$, the corresponding interferometer is an on-axis one, having all cells of each phase shift grids on top of the others.
In this case, we get $\text{expUV}=1$ and $\text{expPQ}= e^{j\pi P\Delta} + e^{j\pi Q\Delta}$, leading to the chromatic term : $\Sum=\Sum(e^{j\pi P\Delta} + e^{j\pi Q\Delta})$.
 As discussed earlier its Taylor developpement is given by :
\begin{equation}
\Sum = 
\sum_{k=1}^N (m_k^0 - n_k^0) +
j\phi\sum_{k=1}^N  (m_k^1 - n_k^1) +
(j\phi)^2\sum_{k=1}^N  (m_k^2 - n_k^2) + \dots
\end{equation}
where $\phi=\pi\Delta$.

The device can be made achromatic up to order $O$ if we simultaneously solve the above system of diophantine equations :
\begin{align*}
\tsum_{k=1}^N m_k &= \tsum_{k=1}^N n_k\\
\tsum_{k=1}^N m_k^2&= \tsum_{k=1}^N n_k^2\\
\dots\dots\dots&\dots\dots\dots\dots\\
\tsum_{k=1}^N m_k^O&= \tsum_{k=1}^N n_k^O
\end{align*}
This system is called a multigrad system and is equivalent to the abbreviated notation : 
$\multi{O}{m_k,k=1,N}{n_k,k=1,N}$ or simply $\multi{O}{m_k}{n_k}$ if there is no ambiguity.
Two sets of integers, solutions of a multigrad system, are called \emph{multigrade} sets, for example we have  $\multi{3}{-11,-3,3,11}{-9,-7,7,9}$.
The problem of solving a multigrad system corresponds to the classical Prouhet-Tarry-Escott problem.
Our problem is a particular multigrad system since the two sets \set{m_k} and \set{n_k} must have the same size and, more important, the $m_k$ must be even and the $n_k$ must be odd.

The solution of our problem hinges around the propertie that if \multi{O}{\alpha_k}{\beta_k} then \multi{O+1}{\alpha_k,\beta_k+c}{\beta_k,\alpha_k+c} for any $c$.
(These properties can be easily proved by induction using the fact that the highest power coefficient of a polynom in $x$ is invariant under the translation $x+c$.) 
Starting from \multi{1}{-1,1}{0,0}, we conserve parity by setting $c=1$ and we get \multi{2}{-1,1,1,1}{0,0,0,2}.
For the next step, we regain symmetry by setting $c=-1$ and get now \multi{3}{-1,-1,-1,-1,1,1,1,1}{-2,0,0,0,0,0,0,2}.
What we use in fact is \multi{1}{-1,1}{0,0} and the property that if
\multi{O}{m_k}{n_k} then
\multi{O+2}{n_k-1,m_k,m_k,n_k+1}{m_k-1,n_k,n_k,m_k+1}, the parity
being preserved.

\section{Second order: distributing the cells on the pupil}

In the preceding section, we built a nulling interferometer with the even cells and the odd cells one on top of the others.
The practical realization of such an interferometer is conceivable, but is far too complicated and will not further be considered here.
The cells must therefore be distributed spatially on the even and odd pupils.

Take for example of the two $4\times 4$ grids ($n_0=4$).
The Prouhet-Pascal solution tells us to distribute the two sets \set{-1,-1,-1,-1,1,1,1,1} and \set{-2,0,0,0,0,0,0,2} on the pupils, but does not tell us where.
To gain some insights on this difficult problem, it is necessary to scrutinize the expUV matrix a step further and look at its expansion.
We have:
\begin{equation}
\text{expUV} = \sum_{k=0}^\infty \frac{1}{k!}j^k\Bigl(\frac{\pi d}{n_0}\Bigr) [\text{UV}]^k\,,
\label{eqa6}
\end{equation}
with (always for the example  $n_0=4$ case):
\begin{equation}
 [\text{UV}]^k = 
\begin{bmatrix}
(-3u+3v)^k & (-u+3v)^k & (u+3v)^k & (3u+3v)^k\\
(-3u+v)^k & (-u+v)^k & (u+v)^k & (3u+v)^k\\
(-3u-v)^k & (-u-v)^k & (u-v)^k & (3u-v)^k\\
(-3u-3v)^k & (-u-3v)^k & (u-3v)^k & (3u-3v)^k\\
\end{bmatrix}\,.
\end{equation}
We observe that the even order matrices $[\text{UV}]^{2p}$ are axisymetric while the $[\text{UV}]^{2p+1}$ odd ones are anti-axisymetric.
Therefore it easy to nullify all the contributions from the odd ones in \eqref{eqa1}, it suffice to distribute the cells in an axisymetric manner.
This cancels out the first order term in \eqref{eqa6} and all the odd ones but, of course, the even ones are still contributing.
{  Next, we observe that the  $[\text{UV}]^k$ matrices are 2-D polynomials in $u, v$ coordinates, therefore they will be cancelled out by a 2-D finite differential operator.}
Following these lines we start with the $2\times 2$ matrices corresponding to the doubled \multi{1}{-1,1}{0,0} solution:
\[
P =
\begin{bmatrix}
0&0\\
0&0
\end{bmatrix}\,,
\quad
Q =
\begin{bmatrix}
-1&+1\\
+1&-1
\end{bmatrix}\,.
\]
{  Indeed, $\text{expPQ} \approx P-Q$ is a differential (gradient) operator.  We obtain the next matrix by using the \multi{O+2}{n_k-1,m_k,m_k,n_k+1}{m_k-1,n_k,n_k,m_k+1} property.
Since there were still some degrees of freedom left, we found the best arrangement keeping the $P-Q$ differential operator property in order to minimize the star/planet ratio of the residual light.}
We got :
\[
P =
\begin{bmatrix}
+2 & 0 & 0 & -2\\
0 & 0 & 0 & 0\\
0 & 0 & 0 & 0\\
-2& 0 & 0 & +2
\end{bmatrix}\,,
\quad
Q =
\begin{bmatrix}
+1 & +1 & -1 & -1\\
+1 & -1 & +1 & -1\\
-1 & +1 & -1 & +1\\
-1 & -1 & +1 & +1
\end{bmatrix}\,.
\]
The algorithm of the third method proceeds following these lines from a $n_0\times n_0$ matrices to the next $2n_0\times 2n_0$ one.
{  This algorithm is very efficient because it produces at last very low star/planet ratio.}
Moreover it possess the desirable property to produces axisymmetrical images which helps finding the location of the planet.
   
\section*{Acknowledgments}
The authors thanks Marie Ygouf for  producing  Fig. \ref{fig:result_model}. This work also received the support of PHASE, the high angular resolution partnership between ONERA, Observatoire de Paris, CNRS and University Denis Diderot Paris 7.
\bibliography{Acronymes,HRA,HRA-a-DRouan,HRA-p-DRouan}
\bibliographystyle{aa}

\end{document}